\documentclass{aipproc}

\usepackage{graphics}
\usepackage{color}

\usepackage{longtable}

\begin{document}

\title{What Can GLAST Say About the Origin of Cosmic Rays in Other 
Galaxies?}

\author{Seth W. Digel\affiliationmark{a,b},
Igor V. Moskalenko\affiliationmark{a,c},
Jonathan F. Ormes\affiliationmark{a},
\\P. Sreekumar\affiliationmark{d}, and
P. Roger Williamson\affiliationmark{e},\\on behalf of the GLAST 
collaboration}
\affiliation{\affiliationmark{a}NASA/Goddard Space Flight Center, 
Greenbelt, Maryland,
\affiliationmark{b}Universities Space Research Association,\\
\affiliationmark{c}National Research Council and Institute for Nuclear 
Physics, Moscow State University,\\
\affiliationmark{d}Indian Space Research Organization,
\affiliationmark{e}Stanford University}

\begin{abstract}
Gamma rays in the band from 20 MeV to 300 GeV, used in
combination with data from radio and X-ray bands, provide a powerful tool
for studying the origin of cosmic rays in our sister galaxies Andromeda
and the Magellanic Clouds.  Gamma-ray Large Area Space Telescope (GLAST)
will spatially resolve these galaxies and measure the spectrum and
intensity of diffuse gamma radiation from the collisions of cosmic rays
with gas and dust in them. Observations of Andromeda will give an external
perspective on a spiral galaxy like the Milky Way.  Observations of the
Magellanic Clouds will permit a study of cosmic rays in dwarf irregular
galaxies, where the confinement is certainly different and the massive
star formation rate is much greater.
\end{abstract}

\maketitle

\section{Introduction}

High-energy gamma rays are produced in interactions of high-energy 
cosmic rays with interstellar matter and photons.  From the resulting 
diffuse emission of gamma rays, the properties of the cosmic rays can 
be inferred (e.g., \cite{hunter97}).  Gamma rays have proven to be a 
useful probe of cosmic rays in the Milky Way, but gamma-ray telescopes to date 
have lacked the sensitivity and angular resolution to permit the same
kind of detailed study of cosmic rays in external galaxies.

\begin{figure}[!t]
\includegraphics[scale=0.42]{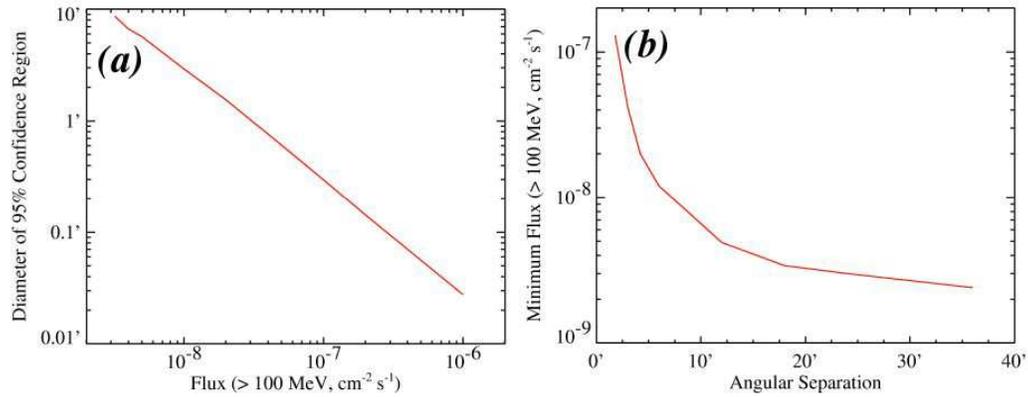} 

\caption{Expected performance of GLAST for localizing and resolving 
point sources.  ($a$) Source localization at high latitudes.  The position
uncertainties for the brightest sources likely will be limited to 
10--20$^{\prime\prime}$ by uncertainty in spacecraft pointing and instrument
alignment.  ($b$)
Minimum flux required to resolve two closely-spaced sources of
equal flux.  For both figures, the sources are assumed to have 
$E^{-2}$ photon
spectra and to be observed at high latitudes in a one-year sky survey.
\label{fig1}}

\end{figure}

      The Gamma-ray Large Area Space Telescope (GLAST) is the next
generation high-energy (20 MeV--300 GeV) gamma-ray astronomy mission.  It
is part of the strategic plan of NASA's Office of Space Science and is
currently planned for launch in 2005.  GLAST will have a factor of 30
greater sensitivity than the Energetic Gamma-Ray Experiment Telescope
(EGRET), launched in 1991 on the Compton Gamma-Ray Observatory.  Derived
performance parameters for our proposed design for the GLAST instrument,
which was selected by NASA in February 2000, are presented in
Table~\ref{table1} and Figure~\ref{fig1}.  See the companion paper by
Ormes {\it et al.} for information about the design and instrumental
response of GLAST, and the Web site
 http://glast.gsfc.nasa.gov for information about the mission.

\begin{table}
\begin{tabular}{lll}
\hline
\tablehead{1}{r}{b}{ }
&\tablehead{1}{c}{b}{EGRET}
&\tablehead{1}{c}{b}{GLAST} \\
\hline
Energy Range &	0.02--30 GeV &	0.02--300 GeV \\
Field of View &	0.5 sr &	2.4 sr \\
Peak Eff. Area	& 1500 cm$^{2}$ &	13,000 cm$^{2}$ \\
Point Source\\Sensitivity\tablenote{Sensitivity at high
latitude after a 2-year survey for a 5-$\sigma$ detection,
units 10$^{-8}$ cm$^{-2}$ s$^{-1}$, for $E > 100$ MeV.} &
    	5 &
    	0.16 \\
Source Location\tablenote{Diameter of 95\% confidence region; range:
    bright sources to sources of flux 10$^{-8}$ cm$^{-2}$ s$^{-1}$ 
    ($E > 100$ MeV).} &	
    5$^{\prime}$ -- 90$^{\prime}$	& 0.2$^{\prime}$ -- 1$^{\prime}$ \\
Mission Life & &		5 years\\
 & & (10-year goal) \\
\hline
\end{tabular}
\caption{Selected Parameters for GLAST and EGRET}
\label{table1}
\end{table}

\begin{figure}[!th]
\includegraphics[scale=0.60]{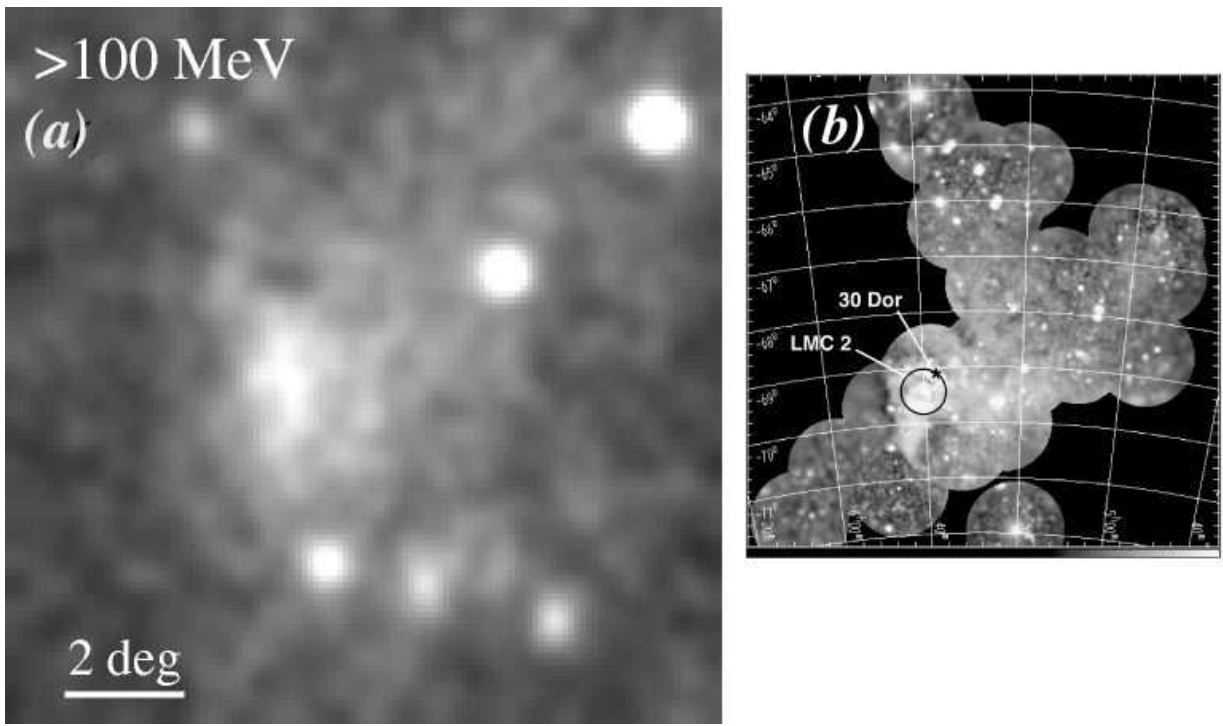} 
\caption{$(a)$ Simulated map of the LMC in gamma rays (> 100 MeV) from a
two-year sky survey with GLAST.  The simulation is based on a model of the
LMC by Sreekumar \cite{kumar99} and also includes foreground diffuse emission from
the Milky Way and an isotropic background consisting of a distribution of
faint point sources.
$(b)$ The LMC in $3/4$-keV X-rays, from a mosaic of pointed
observations with ROSAT \cite{snowden94}.  The intense emission
regions of 30 Doradus and LMC superbubble 2 are indicated.
\label{fig2}}
\end{figure} 

\section{Advances with GLAST}
The only external galaxy that EGRET detected in the light of its
interstellar gamma-ray emission was the Large Magellanic Cloud (LMC),
which was not spatially resolved \cite{kumar92}.  GLAST will be
able to map the diffuse gamma-ray emision of the LMC, as well as the
fainter Small Magellanic Cloud (SMC) and Andromeda (M31) galaxies.

\begin{figure}[t]
\includegraphics[scale=0.64]{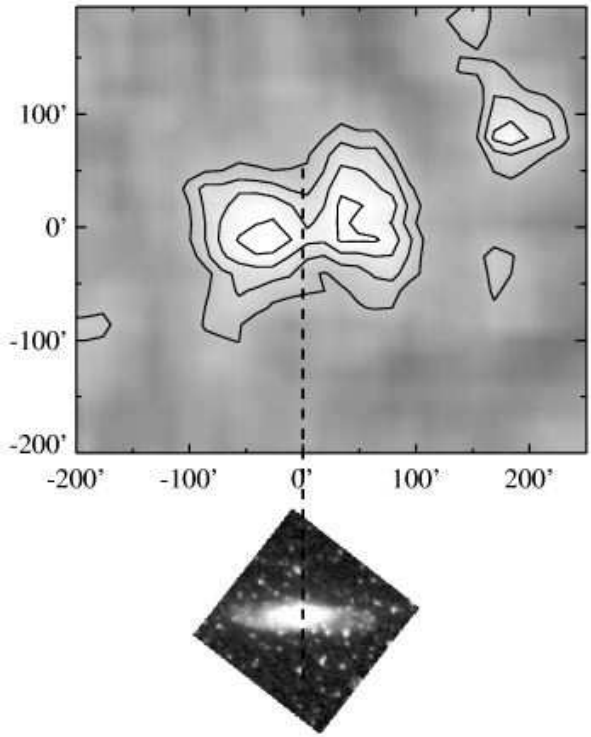} 
\caption{Simulated map of M31 from a five-year sky survey with GLAST 
(approximately equivalent to 6 months of observations with M31 
within $30^{\circ}$ of the center of GLAST's field of view).
The image shows gamma-rays with energies >1 GeV, and has been smoothed to
reduce statistical fluctuations.  The simulated point source in the upper
right indicates the angular resolution of the image, and the inset shows
the location and extent of the optical disk of the galaxy.  The diffuse
emission was modelled based on the distribution of gas in M31, which
extends much further than the optical disk, and the EGRET upper limit for
the galaxy \cite{blom99}.  Contours are spaced 
by $2 \times 10^{-7}$ cm$^{-2}$ s$^{-1}$
sr$^{-1}$ from $2.2 \times 10^{-6}$ cm$^{-2}$ s$^{-2}$ sr$^{-1}$.
\label{fig3}}
\end{figure}

\subsubsection{LMC}
The LMC will be well-resolved by GLAST (Fig.~\ref{fig2}).  The cosmic-ray
distribution can be studied in detail by analyzing the gamma-ray data
together with 21-cm H I and 2.6-mm CO surveys of the interstellar medium
of the galaxy (see, e.g., \cite{cohen88}).  GLAST data should reveal
the degree of enhancement of cosmic-ray density in the vicinity of the
massive star-forming region 30 Dor and the associated superbubble LMC 2
\cite{meaburn80}; see the X-ray image in Fig.~\ref{fig2}.  The LMC 2 superbubble is
among the largest of several prominent superbubbles in the LMC, most of
which represent regions like 30 Dor that are further evolved.  If
superbubbles are sources of cosmic rays distinct from individual
supernovas (e.g., \cite{higdon98}), superbubble 2, which subtends 
$1^{\circ}$,
may be marginally spatially resolved by GLAST.
The extent to which the diffuse emission of the LMC can be attributed to
unresolved gamma-ray pulsars has been considered by Hartmann {\it et al.} 
\cite{hartmann93}
and Zhang \& Cheng \cite{zhang98}.  The expectation is that the pulsar contribution
could be significant, up to ~35\%, for energies $>1$ GeV.  GLAST is unlikely
to detect individual pulsars in the LMC, but may be able to address the
question of the pulsar fraction with a sensitive measurement of the
high-energy spectrum.

\subsubsection{SMC}
Detection of the diffuse gamma-ray flux of the SMC by GLAST will be
useful to verify conclusions about the galactic origin of cosmic rays
(e.g., Sreekumar {\it et al.} \cite{kumar93} based on EGRET 
data).  The non-detection of the
SMC, with a 2-$\sigma$ upper limit of
$5 \times 10^{-8}$ cm$^{-2}$ s$^{-1}$ (>100 MeV) led Sreekumar
{\it et al.} to conclude that the most likely model for the distribution
of cosmic rays in the SMC is one for which the galaxy is disintegrating
and cosmic rays are only very poorly confined.  In this circumstance, the
predicted flux is $(2 \pm 3) \times 10^{-8}$ cm$^{-2}$ s$^{-1}$ (>100 
MeV) \cite{kumar91}, well within the reach of GLAST.
  
\begin{figure}[t]
\includegraphics[scale=0.50]{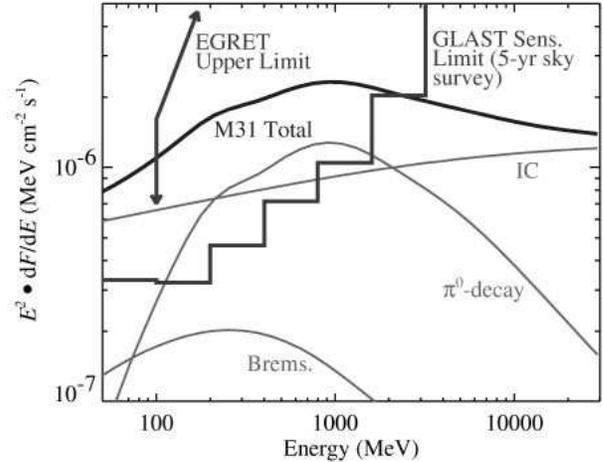} 
\caption{Simulated spectrum of M31, obtained by scaling the
luminosity spectrum of the whole Milky Way derived by
Strong {\it et al.} \cite{strong98} (4-kpc halo model) to match the upper
limit flux of 
Blom {\it et al.} \cite{blom99} for
M31.  The differential flux sensitivity of GLAST for a
five-year sky survey and the EGRET upper limit of Blom {\it et al.}
are also shown, along with the
individual components of the overall spectrum:  inverse
Compton, Bremsstrahlung, and $\pi^{0}$-decay.
\label{fig4}}
\end{figure}

\subsubsection{M31}
The EGRET 2-$\sigma$ upper limit for the gamma-ray flux of M31 is 
$1.6 \times
10^{-8}$ cm$^{-2}$ s$^{-1}$ (>100 MeV), which is much less than the flux of the Milky
Way at M31 \cite{blom99}.  The cosmic-ray densities in M31 are
certainly lower than in the Milky Way, and it has less ongoing massive
star formation.  At a flux level of $1 \times 10^{-8}$ cm$^{-2}$ s$^{-1}$ (>100 MeV), GLAST
will resolve the diffuse gamma-ray emission along the major axis of M31,
to provide information about the relationship between cosmic rays, star
formation rate, and interstellar gas on a large scale (Fig.~\ref{fig3}).  GLAST may
also measure the distribution of cosmic rays in the halo of M31.  Spectral
measurements may allow a global assessment of inverse-Compton,
electron-Bremsstrahlung, and $\pi^{0}$ decay contributions to the interstellar
emission (Fig.~\ref{fig4}).  The unexplained ``GeV excess'' for the Milky Way 
\cite{hunter97} will also be detected if present in M31.  From the gamma-ray
spectra of the Milky Way and M31 the contribution of normal galaxies to
the extragalactic gamma-ray background can begin to be assessed.

Detailed studies of M31 that will be possible with GLAST will benefit from
further development of cosmic-ray models for the Milky Way, for which results
from gamma-ray observations can be checked with direct observations of cosmic
rays.

\section{Conclusions}
For the first time, GLAST will enable spatial and spectral studies of
diffuse gamma rays from external galaxies.  Diffuse, high-energy gamma rays 
are diagnostic of cosmic-ray densities, which especially for the 
proton component are difficult to determine 
from observations at other wavelengths.  When considered together
with X-ray  and radio observations, GLAST data promise a fairly complete 
understanding of the production, propagation, and confinement of 
cosmic rays in Local Group galaxies.

IVM acknowledges support from an NAS/NRC Senior Associateship.

\bibliographystyle{aipproc}
\bibliography{glast2}

\end{document}